\documentclass[twoside,11pt]{article}

\usepackage{jmlr2e}
\usepackage{booktabs} 
\graphicspath{{resources/}} 
\DeclareGraphicsExtensions{.pdf,.jpeg,.png} 
\usepackage[ruled]{algorithm2e} 
\usepackage{todonotes}

\ShortHeadings{EvalNE: A Framework for Evaluating Network Embeddings on Link Prediction}{Mara, Lijffijt and De Bie}
\firstpageno{1}

\begin{document}
	
	\title{EvalNE: A Framework for Evaluating\\Network Embeddings on Link Prediction}
	
	\author{\name Alexandru Mara
		\email alexandru.mara(at)ugent.be \\
		\name Jefrey Lijffijt 
		\email jefrey.lijffijt(at)ugent.be \\
		\name Tijl De Bie 
		\email tijl.debie(at)ugent.be \\
		\addr IDLab, Dept. of Electronics and Information Systems \\
		Ghent University \\
		9000 Ghent, Belgium}
	
	\editor{Unknown}
	
	\maketitle
	
	
	\begin{abstract}%
		In this paper we present EvalNE, a Python toolbox for evaluating network embedding methods on link prediction tasks. Link prediction is one of the most popular choices for evaluating the quality of network embeddings. However, the complexity of this task requires a carefully designed evaluation pipeline in order to provide consistent, reproducible and comparable results. EvalNE simplifies this process by providing automation and abstraction of tasks such as hyper-parameter tuning and model validation, edge sampling and negative edge sampling, computation of edge embeddings from node embeddings, and evaluation metrics. The toolbox allows for the evaluation of any off-the-shelf embedding method without the need to write extra code. Moreover, it can also be used for evaluating any other link prediction method, and integrates several link prediction heuristics as baselines. 
	\end{abstract}
	
	\begin{keywords}
		Network Embedding, Link Prediction, Evaluation, Edge Sampling, Graphs
	\end{keywords}
	
\section{Introduction}\label{sec_intro}

Network Embedding (NE) methods aim at learning low-dimensional representations of network nodes as vectors, typically in Euclidean space. The quality of the resulting representations is assessed through a variety of downstream prediction tasks, with link prediction (LP) one of the most common choices  \citep[e.g.,][]{grover2016node2vec,PRUNE2017,Gao2018bine,cne2019}. LP amounts to estimating the likelihood of the existence of edges between pairs of nodes that are not connected in the input network.
Such edges may be missing from the input network because of incomplete information,
or sometimes because the input network is an earlier snapshot of an evolving network.

The evaluation of LP methods is particularly challenging as it requires a number of additional steps and design choices which can confound the results, are prone to errors, and can harm reproducibility.
First, one needs an (incomplete) training network along with a (more) complete version of that network for testing. It is relatively straightforward to define such training and testing networks whenever different snapshots of a dynamic network are available. For static networks, however, much research has been devoted to determining the best approach to generate these training networks \citep{lpFairEval2002, Yang2015elp, Garcia2015}. In addition to train and test edges, also sets of train and test \emph{non-edges} (also referred to as \emph{negative samples}) are required for evaluating LP. Non-edges are pairs of nodes that are not connected in the input network. The relative sizes of these train and test edge sets as well as the sizes of the non-edge sets are user-defined parameters which vary between scientific works.
Moreover, a challenge specific for NE-based methods for LP is that they usually only provide node embeddings, while the binary classifiers used to predict links require edge embeddings as their input. There are several approaches for deriving edge embeddings from node embeddings \citep{chen2018tutorialne}, the choice of which has a strong impact on the performance of different NE methods \citep{grover2016node2vec}. Also the metrics used to report the accuracy of different methods vary substantially, from AUC-ROC \citep{cne2019} to precision and recall \citep{Wei2017nrcl} or precision@k \citep{wang2016sdne}. Finally, it is not uncommon in recent literature to use recommended default settings for existing methods used as baselines, while tuning the hyper-parameters for the method being introduced. 

In this paper we propose EvalNE, a Python toolbox that addresses one of the critical issues in the evaluation of NE methods and in the field of machine learning in general: the reproducibility of results. The library can be used to replicate the experimental sections of the majority of papers evaluating NE methods on LP for both directed and undirected networks. A command line interface in combination with a configuration file allows users to evaluate any NE method on the LP task, without the need of new software code. This simplifies their model evaluation pipeline, and thus minimizes the risk of errors. Alternatively, EvalNE can be used as an API that provides implementations of most of the building blocks required for LP evaluation, from generating train/test splits to computing evaluation metrics. 

We note that, while a significant part of the functionality of EvalNE is geared towards NE methods, it is capable of evaluating also other kinds of LP methods.

\section{Architecture}\label{sec_archit}

\begin{figure}
	\centering
	\includegraphics[width=\textwidth]{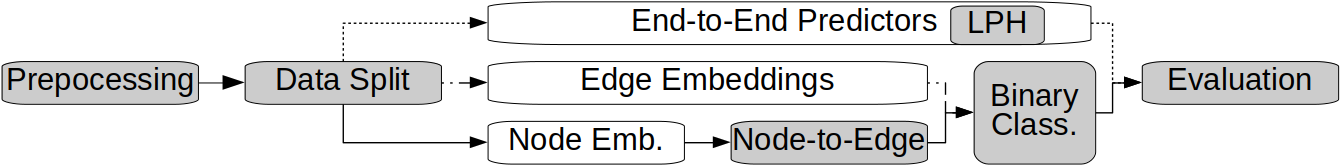}
	\caption{Diagram showing the types of methods that can be evaluated using EvalNE. The gray blocks represent modules provided by the library while the remaining ones are the methods to be evaluated and thus user-specified. 
	The library allows for the evaluation of end-to-end prediction methods (several 
	LP heuristics are included as baselines here), edge embedding methods, and node embedding methods. 
		}
	\label{fig:diagram}
\end{figure}

EvalNE has been designed as a pipeline of interconnected and interchangeable building blocks. This modular structure, presented in Figure~\ref{fig:diagram}, allows for the evaluation of different types of methods and simplifies code maintenance and addition of new features. 
An evaluator class integrates all the building blocks and is the main object the user interacts with. The library also provides different levels of data encapsulation in order to reduce the risk of incorrect model evaluation. Next, we describe the core building blocks of EvalNE.

\paragraph{Preprocessing} The toolbox offers a variery of functions to load and store networks, to prune nodes based on their degrees, remove self-loops, relabel nodes, obtain sets of specific types of edges, restrict networks to their main connected components and obtain common network statistics. These functions can be used in combination with other methods provided by libraries such as NetworkX. 

\paragraph{Data Split} In order to perform LP, sets of train and test edges need to be selected from the input networks. The training set is generally required to span all nodes in the network and induce a training network with a single connected component. 
EvalNE provides an edge set selection algorithm which satisfies these requirements, can generate edge sets of different sizes and is orders of magnitude faster than the naive approach currently used in practice (included in EvalNE for comparison).

The library also offers choice of sampling the sets of \textit{non-edges} required for LP under either the \emph{open world} or the \emph{closed world} assumption. This determines if the set of train \textit{non-edges} does not overlap with the set of train edges (open world) or with both the sets of train and test edges (closed world). The sizes of these sets are tunable parameters.

\paragraph{LP Heuristics} EvalNE contains a set of heuristics which can be used as baselines for the LP task. These methods are: common neighbours, Jaccard coefficient, Adamic-Adar index, resource allocation index, preferential attachment and Katz. Additionally, the library includes a random prediction method. These baselines are also defined for directed networks by constraining the analysis to the \textit{in} or \textit{out} node neighborhoods.

\paragraph{Node to Edge embeddings} Many NE methods only provide embeddings of the network nodes. In order to perform LP, however, edge embeddings are required. In EvalNE we include several methods to compute edge embeddings from node embeddings. The user can select any, or a combination of, the following binary operators: average, hadamard, weighted $L_1$ and weighted $L_2$ \citep{grover2016node2vec}.

\paragraph{Binary Classification} To predict the existence or absence of a link from a set of given edge embeddings, EvalNE uses, by default, regularized logistic regression with $L_2$ penalty and 10-fold cross validation of the regularization parameter. The library, however, is flexible and allows for any other binary classifier (e.g. from Sklearn) to be used. 

\paragraph{Evaluation metrics} EvalNE can evaluate the scalability and accuracy of embedding methods. The scalability is measured trough wall clock time and the LP accuracy using two types of metrics: fixed-threshold metrics and threshold curves.

Fixed-threshold metrics summarize method performance to single values. The framework implements the following: confusion matrix (TP, FN, FP, TN), precision, recall, fallout, miss, accuracy, F-score and AUC-ROC. Threshold curves present the performance of methods for a range of threshold values. EvalNE provides precision-recall curves \citep{lpFairEval2002} and ROC curves \citep{Fawcett04rocgraphs}. The most suitable metrics based on the evaluation setup are recommended to the user. 

\section{Relation to other software}\label{sec_related}

To the best of our knowledge only two libraries for the evaluation of NE methods currently exist.
OpenNE (\url{github.com/thunlp/OpenNE}) 
is a recently proposed Python toolbox which provides implementations of state-of-the-art embedding methods and evaluation on multi-label classification tasks. GEM \citep{Chen2017oneckalg} is a similar Python framework which also implements several embedding methods and includes basic functionality for evaluating multi-label classification, visualization, and LP. 

However, both these libraries require Python implementations of the NE methods evaluated that comply with pre-defined interfaces. As these implementations are not always feasible or practical, we have designed EvalNE to evaluate any NE method written in any language by delegating the method execution to the system command line interface.
Moreover, EvalNE is the only currently available library which provides full automation of the method evaluation pipeline, parameter tuning, and edge sampling capabilities. Finally, no other open-source libraries include implementations of similar LP heuristics for both directed and undirected networks.

\section{Source Code and Documentation}\label{sec_scanddoc}

The source code and documentation of EvalNE are available on Bitbucket\footnote{EvalNE Bitbucket repository https://bitbucket.org/ghentdatascience/evalne/src/master/} and GitHub\footnote{EvalNE GitHub repository https://github.com/Dru-Mara/EvalNE} and are provided under the MIT free software license. The abovementioned platforms provide different mechanisms for the community to contribute to the project such as bug tracking, feedback submission and pull requests for adding new features. The code style complies with PEP 8 and the documentation follows the numpy docstring format. The toolbox is compatible with Python 2 and Python 3 and can be easily installed using \texttt{pip}. Supported platforms include Linux, Mac OS X, and Microsoft Windows. EvalNE only depends on a small number of popular open-source Python packages, and follows their coding guidelines: NumPy, SciPy, NetworkX, Scikit-learn and Matplotlib. Other packages such as OpenNE or GEM are recommended as they provide implementations of different NE methods.

The toolbox documentation includes instructions on the installation and use both as an API and as a command line tool. Simple examples of the high-level use of the library are also included as well as more advanced examples of the low-level use and integration with existing Python code. Finally, the library contains pre-filled configuration files which reproduce the experimental sections of several influential papers on NE.

\section*{Acknowledgements}

The research leading to these results has received funding from the European Research Council under the European Union's Seventh Framework Programme (FP7/2007-2013) / ERC Grant Agreement no. 615517, from the FWO (project no. G091017N, G0F9816N), and from the European Union's Horizon 2020 research and innovation programme and the FWO under the Marie Sklodowska-Curie Grant Agreement no. 665501.

	\bibliography{bibliography}
	
\end{document}